\documentstyle[11pt,aaspp4]{article}
\slugcomment{To Appear in The Astronomical Journal (August 1999)}

\lefthead{Gizis et al.}
\righthead{Hyades}

\begin{document}

\title{A 2MASS Survey for Brown Dwarfs Towards the Hyades}

\author{John E. Gizis}
\affil{Department of Physics and Astronomy, University of Massachusetts, 
Amherst, MA 01003}
\author{I. Neill Reid\altaffilmark{1}}
\affil{Palomar Observatory, 105-24, California Institute of Technology,
  Pasadena, CA 91125}
\author{David G. Monet}
\affil{U.S. Naval Observatory, P.O. Box 1149, Flagstaff, AZ 86002}

% affiliations, which are identified by the \altaffilmark after each name. 

\altaffiltext{1}{Visiting Research Associate, Observatories of the 
Carnegie Institute of Washington}

\begin{abstract}
We present results of a search for very low mass stars and brown
dwarfs towards the central region of the Hyades cluster.  Using Two 
Micron All Sky Survey
(2MASS) near-infrared and Second Palomar Sky Survey 
(POSSII) optical photometry, we select candidates corresponding
to spectral types M8 through L4 at the distance of the Hyades.
Spectroscopic followup data indicate that none are cluster
members.   The lack of members in the
mass range $0.08 < M < 0.06 M_\odot$ is consistent with
prior evidence that this 600 Myr old cluster is deficient in
very low mass objects.  
Three of our objects are spectral type M8, with
surface gravity sensitive features intermediate between 
dwarfs and giants.  We interpret these objects as background 
brown dwarfs, perhaps associated with the
Taurus star formation region.
\end{abstract}

\keywords{open clusters and associations: individual (Hyades) ---
stars: low-mass, brown dwarfs}

\section{Introduction \label{intro}}

In recent years the discovery of numerous objects at and below 
the hydrogen burning limit has transformed the field of 
brown dwarf studies.  Deciphering the nature of these objects, which include
the coolest M dwarfs, the new L dwarfs (\cite{denis}; \cite{k2m}, 
hereafter K2M),
and the methane (``T-type'') brown dwarf Gl 229B (\cite{gl229b}), 
is simpler when additional information is present.  Brown dwarfs in open
clusters are particularly valuable, as the more massive cluster members
can be used to determine both the age and the composition.  
 
Searches of the Pleiades cluster have led to the discovery
of numerous brown dwarfs (\cite{stauffer1}; \cite{stauffer2}; 
\cite{rmbmzo96}; \cite{pleiades}; 
\cite{bsmbwb98}, hereafter B98).   We present results of a search
for brown dwarfs in the Hyades cluster, which has an age of
625 Myr and whose center lies at 46.3 pc (\cite{hipphyades}).   
In comparison to the Pleiades, the Hyades is both older and more extended 
on the sky, making it a more difficult target for CCD imaging
surveys.  Furthermore, its position in Taurus means that it
is projected against a background of young pre-main sequence stars 
and strong reddening,
both of which complicate the identification of red brown dwarfs on
the basis of color.

Previous searches of the Hyades were not sufficiently sensitive to detect
isolated substellar mass dwarfs.
Searches based upon Palomar Schmidt blue (O, corresponding
roughly to B-band) and red (E, corresponding roughly to R-band)
plates (\cite{luyten}; \cite{r92}) have identified many
M dwarf members.  Over a more limited area, these have been 
extended 
to fainter magnitudes by
selecting objects detected either only on POSS I red (E) plates 
(\cite{bryja}) or IV-N (I-band) plates  (\cite{lh89}).
The coolest objects from these searches are M6.5 (\cite{bryja}) and
M8.5 (\cite{rh99}, hereafter RH99) -- above the stellar/substellar limit
for the Hyades.   Searches for brown dwarf companions have
probed deeper:  Keck K-band imaging (\cite{mac})
and HST I-band imaging (\cite{rg97b}) 
were sensitive to massive brown dwarf secondaries, but
found only a few stellar companions.  The hydrogen burning limit
is expected to be cooler than M8.5, approximately spectral type L0. 

In Section~\ref{sample}, we describe the selection of our sample
from 2MASS and POSSII data, as well as the follow-up spectroscopic
observations.  In Section~\ref{no_brown_dwarfs}, we discuss the
results for the Hyades cluster.  In Section~\ref{background_bd},
we discuss objects that we interpret as background, young
brown dwarfs.  Finally, in Section~\ref{conclusions}, we summarize
our results.

\section{Sample Selection\label{sample}}

\subsection{2MASS and POSSII data}

We use the 2MASS near-infrared data (Skrutskie et al., in prep.) and the  
U.S. Naval Observatory PMM scans (Monet 1999, priv. comm.) 
of the Second Palomar Sky Survey (\cite{poss2}) to search 28 square
degrees at the center of the cluster.  The 2MASS data provide 
JHK$_s$ photometry and positions.  For the data used in this study,
a signal-to-noise of 10 is achieved at a limiting magnitude of 
J$=16.3$, H$=15.4$, and K$_s=14.8$.\footnote{Further 
information on 2MASS is available at http://www.ipac.caltech.edu/2mass/}

\begin{figure}
\plotone{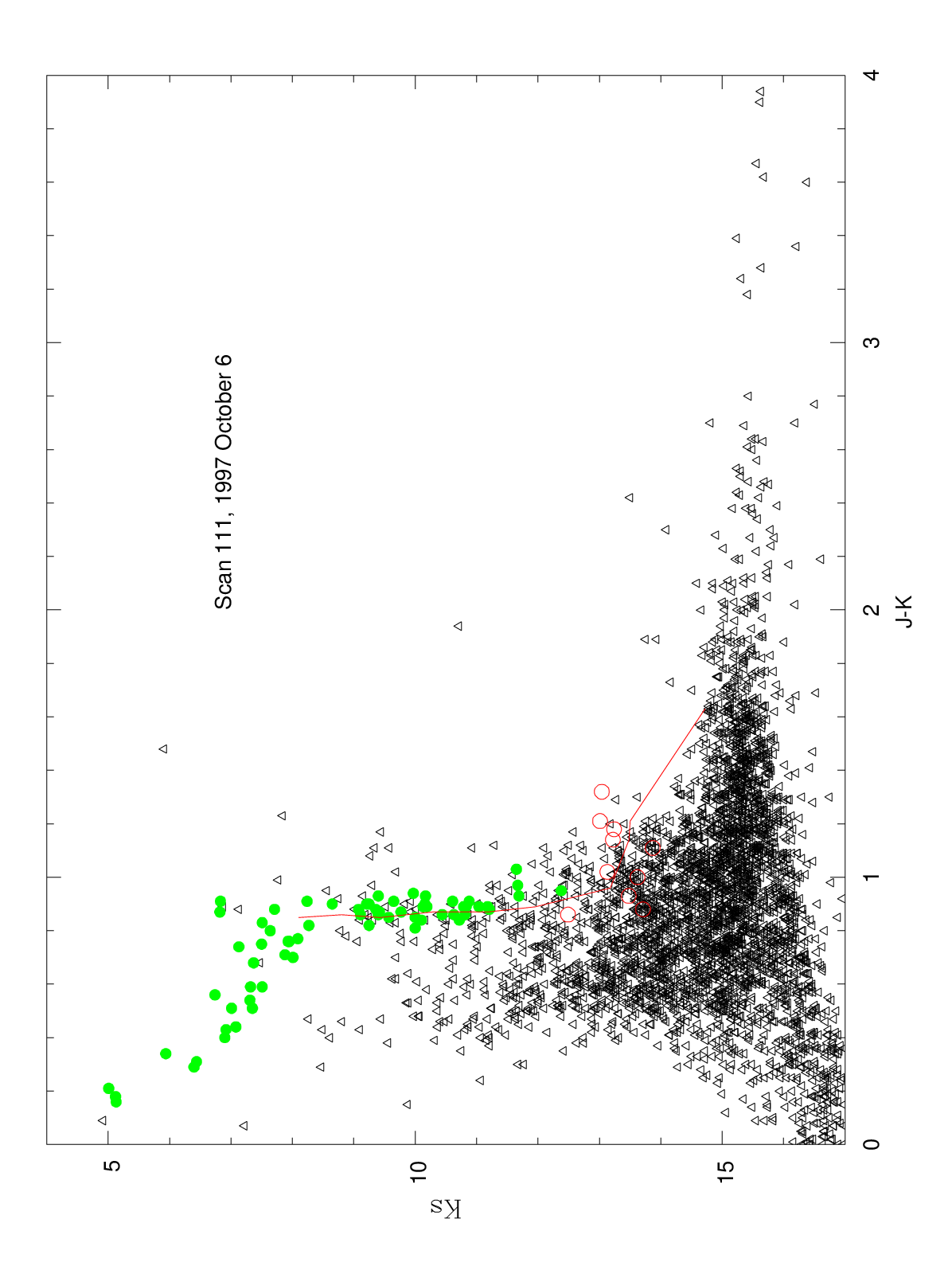}
\caption{ 2MASS color-magnitude diagram.  All objects detected
in this particular scan are shown as open triangles.  2MASS data for
selected known Hyades members (from a larger area) are shown as 
solid circles.  Open circles represent the Leggett \& Hawkins
objects.  The solid line shows the Leggett (1992) M dwarf main sequence
extended to GD 165B. 
\label{fig_jkk}}
\end{figure}

The color-magnitude diagram in the direction of the Hyades is
given in Figure~\ref{fig_jkk} for a limited area (0.87 sq. degrees).  
Also superimposed
are 2MASS photometry for known Hyads and the Leggett \& Hawkins
(1989) objects, as well as a line indicating the position of the
field disk main sequence using the data of Leggett (1992).  
The sequence was extended to the cooler L dwarfs by adding 
the L4 dwarf GD165B.
Selection of Hyades candidates based on 2MASS colors alone would result 
in a sample contaminated extensively by background objects.
Fainter than K$=15$ the photometry
becomes noisy enough that it is useless for identifying candidates.

\begin{figure}
\plotone{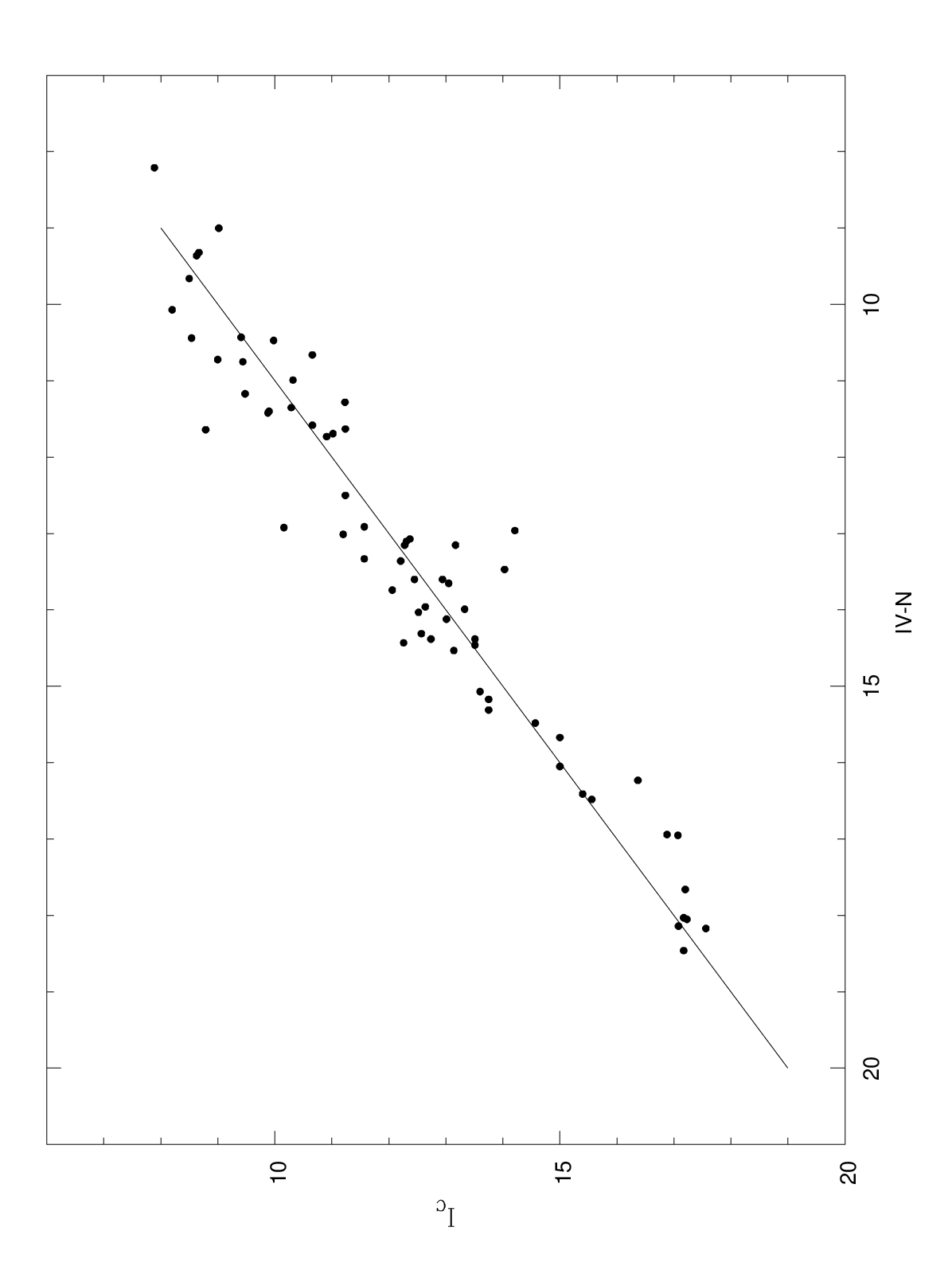}
\caption{ The instrumental POSSII IV-N photometry compared to 
published $I_C$ values for Hyades members.  The relation 
$I_C = (IV-N)-1$ is adequate for our purposes.
\label{fig_in}}
\end{figure}

Most background objects can be eliminated with
the addition of the POSSII photometry.  POSSII provides B (IIIa-J),
R (IIIa-F), and I (IV-N) photometry.  
We have crudely calibrated the photographic instrumental magnitudes
by applying a constant zero-point determined by comparing 
the instrumental magnitudes for known Hyades against 
previously published CCD photometry (\cite{r93}; \cite{lhd94}).
This zero-point calibration is adequate for the identification
of very red candidates, and is illustrated in Figure~\ref{fig_in}.
The scatter about the relation is $\sigma = 0.45$ magnitudes.  
In Table~\ref{table-known}, we list the 
2MASS and POSSII magnitudes for some of the coolest Hyades candidates 
identified by
other surveys.  The POSSII plates reach substantially deeper
than the older POSSI plates.
Bry 804, which was discovered on a red (E) POSSI 
plate but did not appear on the blue (O) plate (\cite{bryja}), is detected
on even the IIIaJ POSSII plate.  Similarly, most of the
Leggett \& Hawkins N-plate only detections are detected on the F plate.  
The completeness of the 2MASS point sources detections is greater than
99\%.

\begin{figure}
\plotone{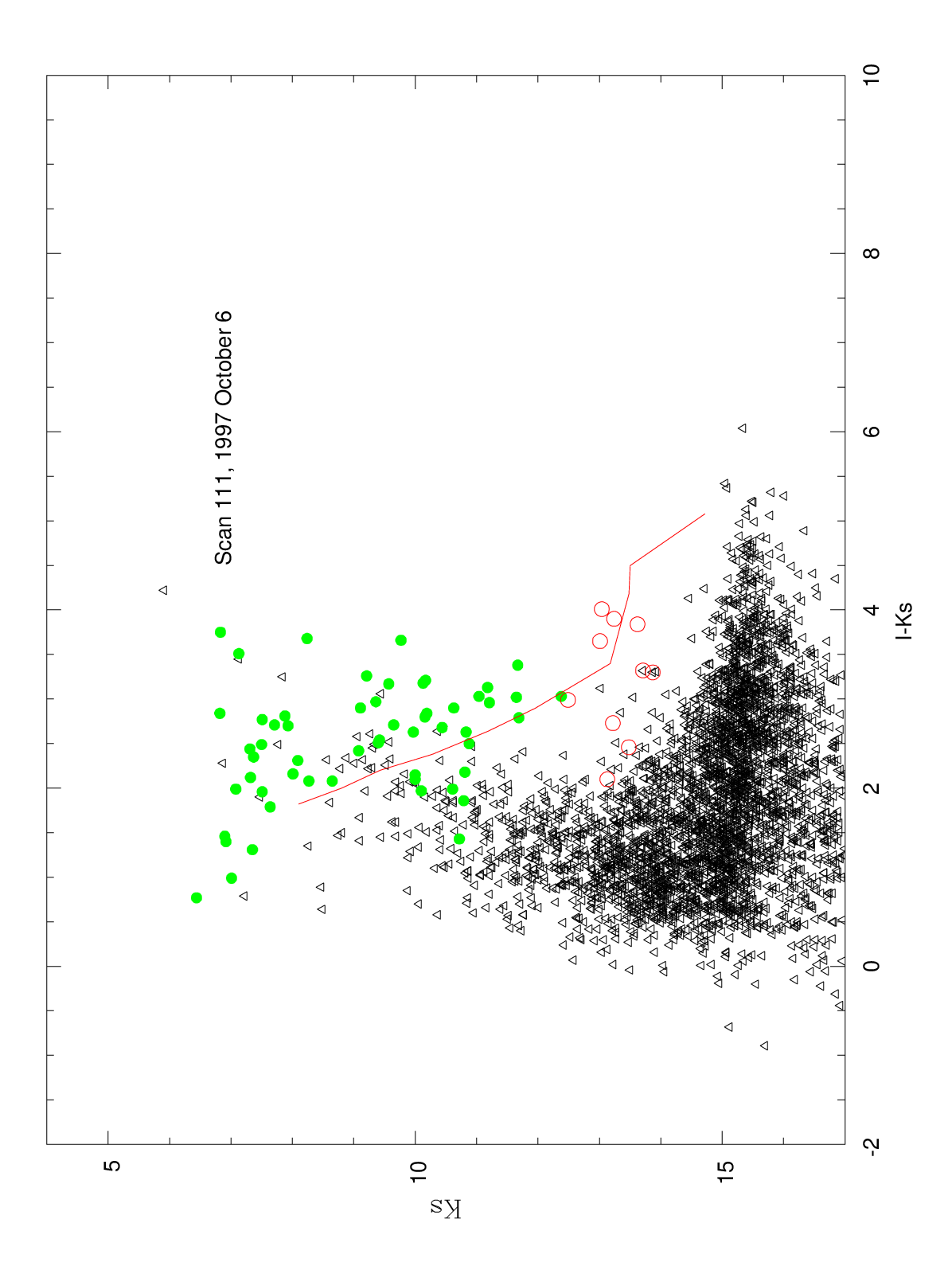}
\figcaption{ 2MASS-POSSII color-magnitude diagram.  The I magnitudes
are instrumental with only a mean zero-point applied.  Systematic
errors are evident at the bright end, but faint, cool Hyads can be
distinguished, such as the coolest of the LH dwarfs. Symbols are as
in Figure~\ref{fig_jkk}.  
\label{fig_ikk}}
\end{figure}

Figure~\ref{fig_ikk} plots the I-K, $M_{K_s}$ diagram for the same
scan as in Figure~\ref{fig_jkk}.  Although there are obvious
distortions in the color-magnitude diagram, 
the expected position of very cool
Hyades members now stands off better from the background objects.  
Note that the LH stars that scatter bluewards into the background have
been shown to be nonmember mid-M dwarfs by RH99.  

We impose the following criteria:
\begin{itemize}
\item Detection in all three 2MASS bands
\item Non-detection on POSSII J plate
\item Non-detection on POSSII F plate or  $R_F-K_s \ge 5.5$
\item Non-detection on POSSII I Plate or $I_N-K_s \ge 2.0$ 
\item $J-Ks \ge 1.00$
\item $12.8 \le K_s \le 14.9$ and $K_s \le 2.53 \times (J-K_s) + 10.9$
\item $J-H > 1.167 \times (H-K_s) - 0.67$
\end{itemize}
The first five requirements ensure a real source that is red in 
both near-infrared and optical-to-near-infrared colors.  
The penultimate requirement is that the objects are no fainter in 
$K_s$ than 0.3 magnitudes below the Hyades main sequence (a necessary
requirement due to the depth of the cluster).
The final requirement eliminates objects with a large infrared
excess, which we expect to be background sources associated with star 
formation regions. 
Finally, each candidate was inspected on the DSS POSSI and POSSII scans
and the 2MASS images to eliminate false sources, generally 
bright star artifacts.  Some apparently real objects whose photometry is
likely to be corrupted by bright star halos were also excluded.
Ideally, proper motions would provide an additional test.  
In practice, the epoch difference between
2MASS and POSSII is insufficient to identify
Hyads with high confidence, while our new candidates are all too faint
to appear on POSSI.

\begin{figure}
%\plotone{fig_pos.ps}
\plotone{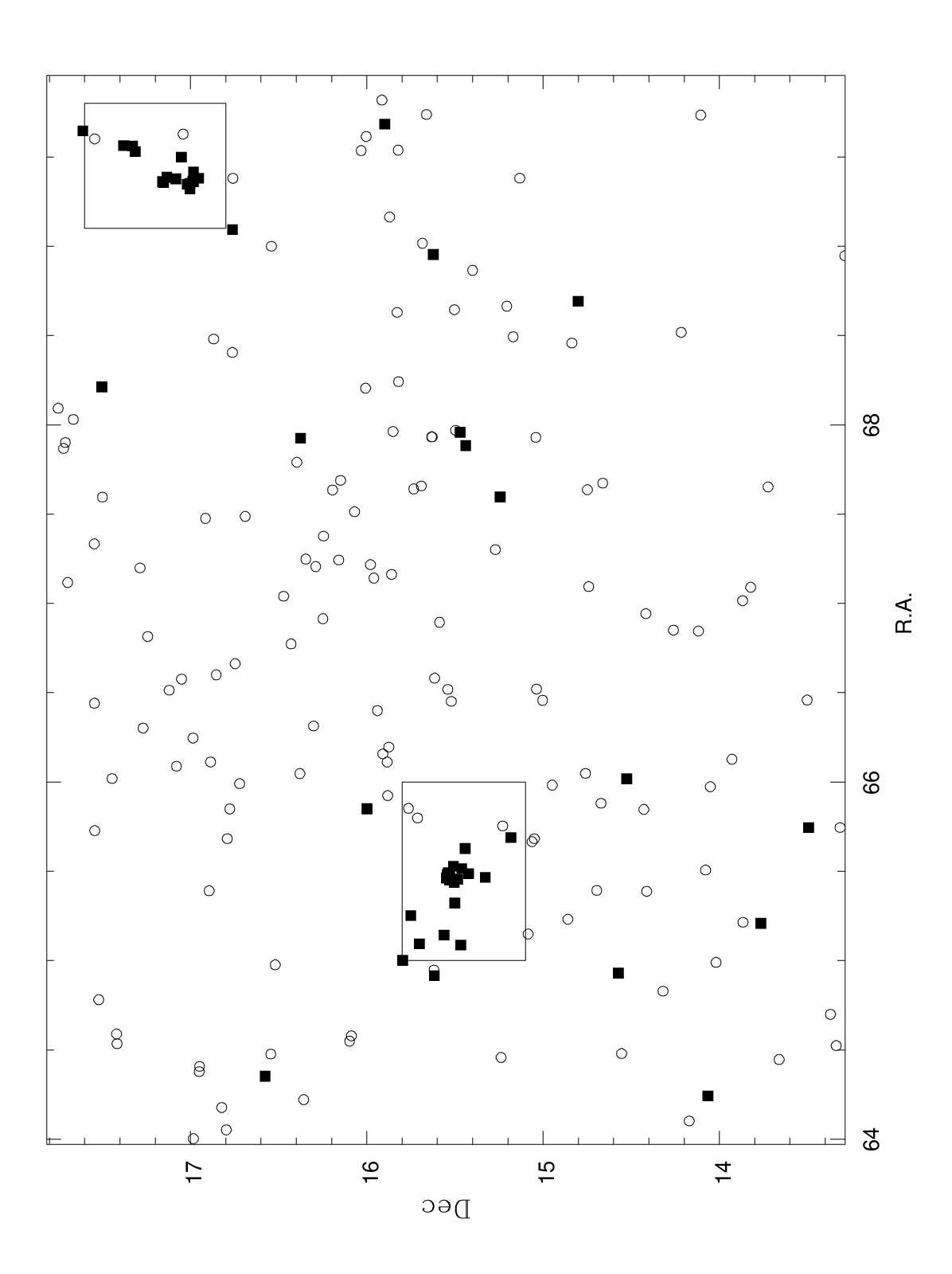}
\caption{\footnotesize Distribution of brown dwarf candidates.  All objects
selected by our photometric criteria are shown as the solid squares.
For comparison, known Hyades members are shown as open circles.  
The cluster is much larger than our survey area, so members
are randomly distributed.  Our candidates, however, show 
evidence for two ``clumps'' (marked as rectangles) which 
we attribute to background clouds or clusters
\label{fig_pos}}
\end{figure}

\begin{figure}
\plotone{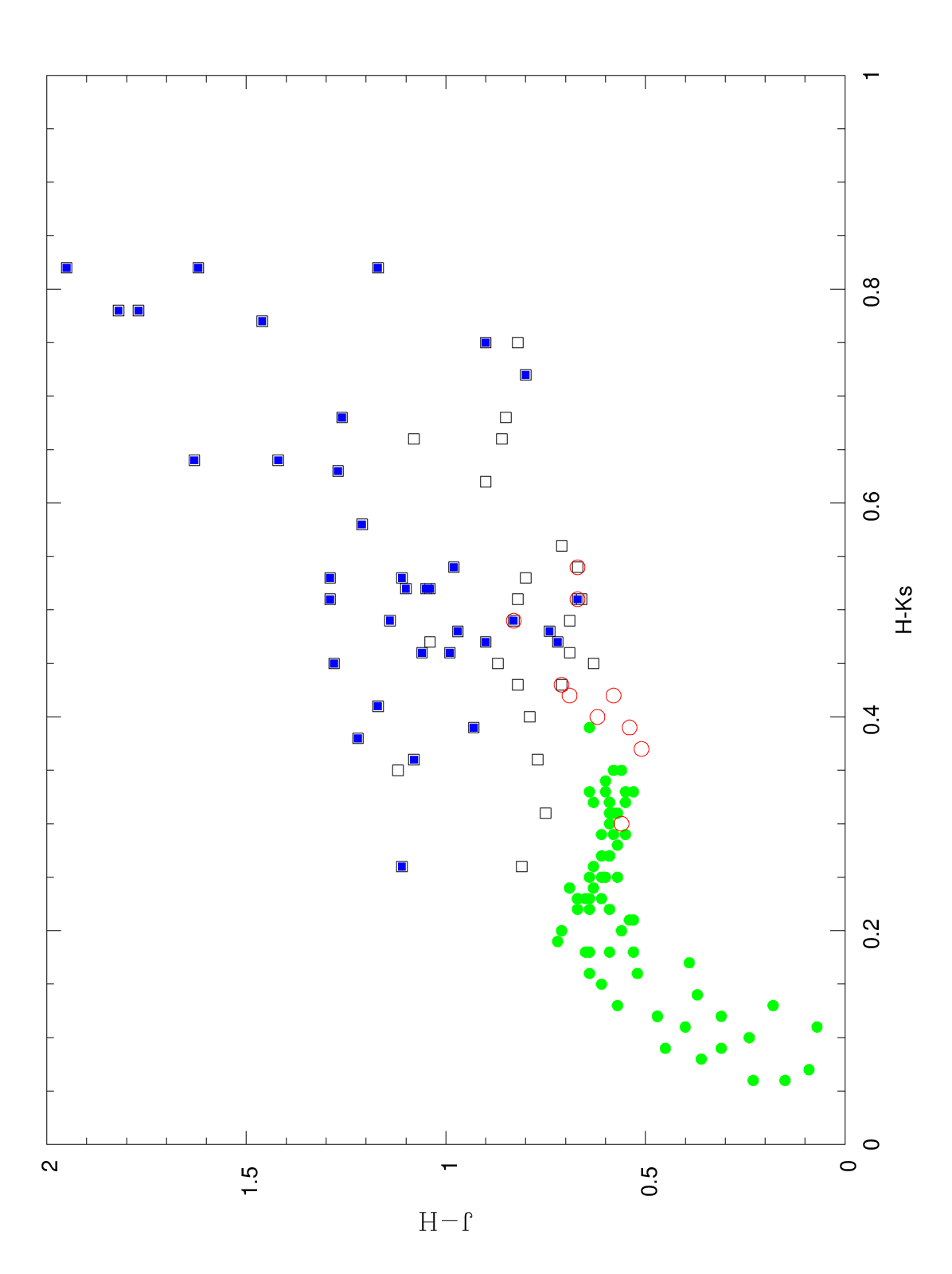}
%\plotone{fig_jhhkpos.ps}
\caption{2MASS color-color diagram.  Known Hyades are 
solid circles, LH objects are open circles, new candidates are
open squares, and ``clump'' objects are solid squares.  
Note that the ``clump''
objects appear to be reddened background  stars.
\label{fig_jhhkpos}}
\end{figure}

After applying the above criteria, we are left with 58 candidates
in the central Hyades region.  Figure~\ref{fig_pos} shows the
positions of both these candidates and the known Hyades members.  
Since the cluster has an observed radius of
at least 10 degrees on the sky, true members are randomly
distributed across our region.  The observed clumps,  at 
$(\alpha,\delta)= 65.5\deg,+15.5\deg$ and 
$(\alpha,\delta)= 69.5\deg,+17.1\deg$ therefore are not associated with
the cluster.  This is supported by Figure~\ref{fig_jhhkpos}, which
plots the near-infrared color-color diagram.  The ``clump'' objects
are consistent with being reddened background main sequence stars,
while the non-clump objects are more likely to lie in the
ultracool M/L dwarf region.  We exclude the 37 ``clump'' objects
from further analysis.  We do {\it not} apply a J-H,H-K$_s$ cut
on the non-clump objects to exclude the background reddened main 
sequence stars because 
a few of the K2M L dwarfs lie in this region of color-color space
(see their Figure 14), and because the uncertainties in the
photometry become significant for the reddest 
candidates.\footnote{It should be noted that two of the K2M L dwarfs have
apparent infrared excesses, and would have been excluded
by our analysis.  The nature of these sources is not yet understood,
but we assume that they are not outliers due to any characteristic
(age, metallicity) in common with the Hyades.  The most likely 
explanation is simply erroneous colors, as both objects have 
sufficiently large uncertainties in color to explain their shift
from the L dwarf sequence.}  
A few of the clump objects have J-H,H-K colors that could be
consistent with M/L dwarfs, perhaps associated with the clumps
themselves; it should be noted that
there may be interesting young objects associated with these clumps.
Table~\ref{table-targets} lists the candidates that pass
these criteria.  Spectra were also obtained of
a number of objects that did not meet our selection criteria
above or were drawn from a larger area.  These objects are
also listed in Table~\ref{table-targets}.

Many of the Leggett \& Hawkins objects are excluded by our
selection criteria -- only LH0416+14 (background M7.5) and 
LH0418+13 (Hyad M8.5) enter our list.  RH99's
data show that all of the others are earlier than M8, so our
exclusion of them is correct.  Four of the objects
listed in Table~\ref{table-known} are excluded because they are
not red enough in J-K$_s$.  Another one, LH0427+12, is just red enough
in J-K$_s$, but is excluded due to its $B_J$ detection.  
LH0418+15 and LH0419+15 lie in the ``clumps,'' and indeed
their J-K$_s$ color is too red for their M6.5-M7 spectral types.
Finally, LH0420+13 is considered too bright at K$_s$ to be a
member.

\section{Spectroscopy\label{spectra}}

Spectroscopy was obtained at the Las Campanas 100-in. and
the W.M. Keck Observatory.  Only the brightest of our candidates
were observable at Las Campanas.  The 100-in. observations used
the Modular Spectrograph with a Tektronix CCD (denoted ``Tek 5'') 
and a 600 l/mm grating blazed at $7500$ \AA.  This yielded
useful spectra over the range $6100 - 9400$ \AA.  
The Keck observations used the same instrumental
setup as in K2M:  LRIS (\cite{lris}) with 
the 400 l/mm grating blazed at 8500 \AA, yielding $9$ \AA resolution
spectra over the wavelength range $6300 - 10100$ \AA.  
We did not reobserve LH0418+13 or LH0416+14, since RH99 observed them.  
The spectral information is listed in Table~\ref{table-targets}.

While our spectra are not of sufficient resolution to measure
radial velocities or detect lithium absorption, the spectra allow 
contaminating background
earlier-type M dwarfs and reddened stars to be rejected.  
As can be seen from the cluster main sequence in Figure~\ref{fig_jkk},
M1-M7 dwarfs all have nearly identical J-K$_s$
colors.  Background M dwarfs can thus be seen in a large
volume behind the Hyades, and a 1 or 2 sigma error in the
color will scatter them into our sample.  While our
POSSII magnitudes help with this effect, the R and I magnitudes
are not very precise.  Our spectra show that only four
of our targets are actually M8-M9 dwarfs, and {\it none} are L dwarfs.
Spectral types were determined by visual comparison to stars on the
Kirkpatrick et al. (1995) system and by the Martin et al. (1995) PC3 index.
All candidates with J-K$_s>1.2$ were observed, with one
exception.  This object has detected on both the F and N 
plates yet has a J-K$_s$=1.52.   
To be conservative we initially retained sources with extremely
red (J-K$_s$) colours which were also detected on the POSS II
F plates. However, our spectroscopy shows that all such objects prove to
be background contaminants, and indeed
the red L dwarfs with similar K$_s$ magnitudes
observed by K2M are not detected on the POSSII F plates.   
We therefore consider this object to be non-member.

Including the two previously known LH objects, 
there are four $>$M7 objects in the Hyades central region
with {\it photometric} properties consistent with membership,
plus an additional object (2MASSW J0428299+235848) which lies
in the outer parts of the cluster where we are incomplete.  
However, only LH0418+13 has {\it spectroscopic} properties 
consistent with membership.  
The remaining four objects (LH 0416+14, and three new 2MASS objects) 
all have spectral features that
distinguish them from ordinary M8-M9 dwarfs.  The most
notable is that all show a weak Na doublet at $8200$ \AA,
suggesting that they have surface gravity intermediate between
dwarfs and giants.  
These three new 2MASS objects are similar to the 
five LH objects discussed by RH99 as young brown dwarf candidates.
We discuss their significance further in Section~\ref{background_bd}.

\section{Still No Brown Dwarfs in the Hyades\label{no_brown_dwarfs}}

Our search was designed to be sensitive to M8 to L4 dwarfs
(i.e., vB10 to GD 165B) at the distance of the Hyades.  
We estimate the corresponding mass range using
the theoretical models of Burrows et al. (1993).
Their Model X has the stellar/substellar limit at
$0.0767 M_\odot$.  For a 600 Myr old cluster, objects at
$M=0.076 M_\odot$ are predicted to have 
$T_{eff}=2502$ and $M_{BOL} = 13.36$.  For comparison,
the Baraffe et al. (1998) $M=0.075 M_\odot$ has
$T_{eff}=2342$ and $M_{BOL}=13.51$ at 637 Myr.  
Tinney et al. (1993)
have measured empirical bolometric corrections for cool
dwarfs and have shown that the $BC_K=3.2$ is appropriate
for M7-M9 dwarfs.  Thus, we estimate that the stellar/substellar
limit lies at $M_K =10.3$, or $K=13.6$. 
The upper limit of the survey corresponds to the
coolest known Hyades candidate, LH0418+13, which
RH99 estimate to have $M_{BOL} =12.78$ and
$M \approx 0.083 M_\odot$ using Model X.  
A $0.06 M_\odot$ object is predicted to have $T_{eff} = 1900 K$
and $M_{BOL} = 14.69$.  This is approximately the temperature
of the L4 dwarf GD 165B (K2M; \cite{k99a}).  Applying
Tinney et al.'s $BC_K = 3.3$ leads to a prediction of $M_K = 11.4$
or $K=14.7$.  Our search could therefore
detect objects of about $0.085 M_\odot > M > 0.06 M_\odot$,
but only the already known LH0418+13 at the uppermost end
of this range was found.

\begin{figure}
\plotone{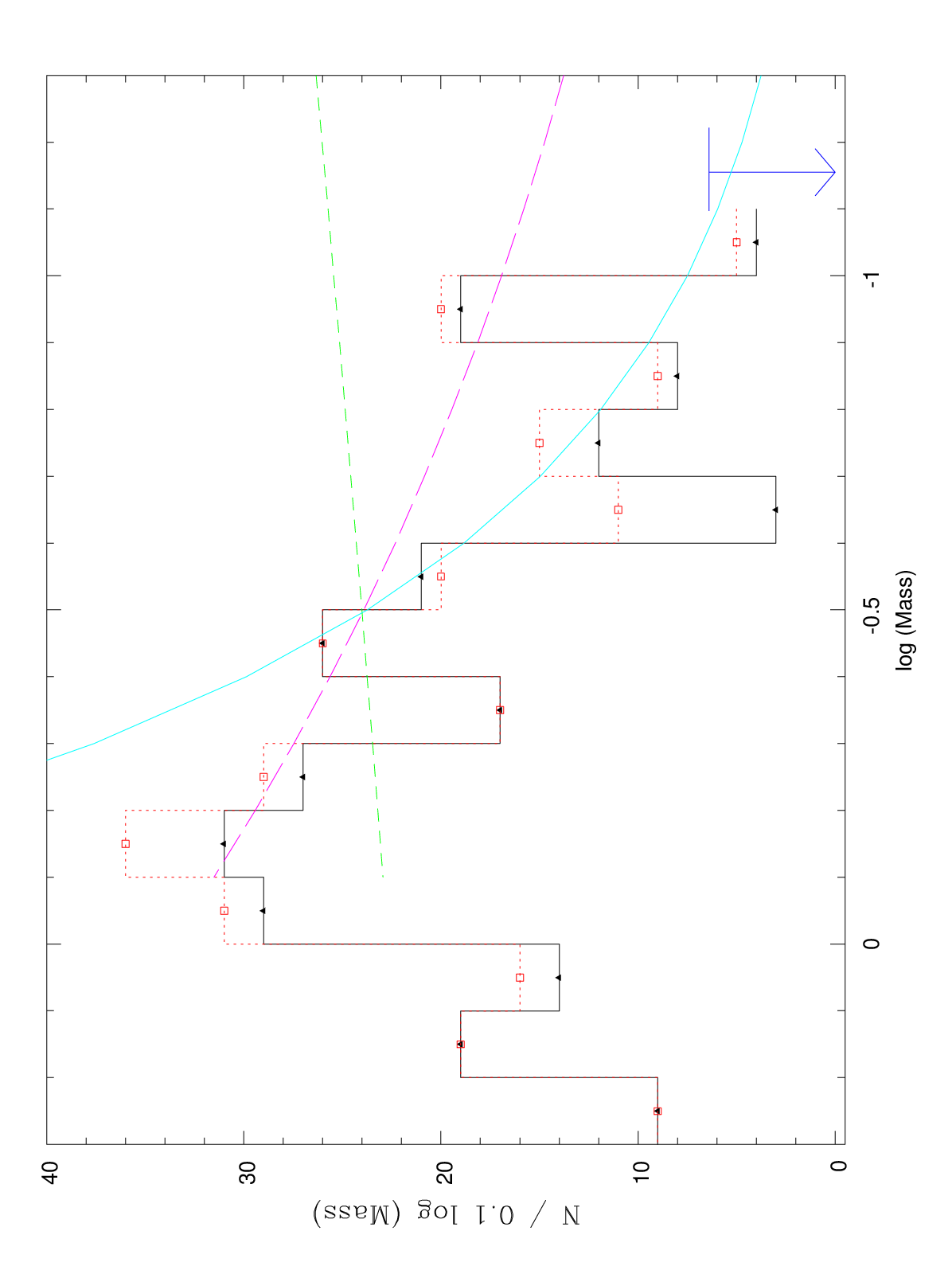}
%\plotone{fig_mf.ps}
\caption{\footnotesize Mass function of the Hyades as computed by Reid \&
Hawley (1999), with our $90\%$ confidence upper limit shown.
The solid histogram plots the mass function of systems, while 
the (higher) dotted histogram adds the known resolved secondaries.
Our limit continues the trend already evident below $0.25 M_\odot$ ---
the number of objects per log mass is steeply declining.   
Also shown are the power law mass functions with 
$\alpha=1.05$ (short dashes), 0.7 ((long dashes), and 0.0 (solid).
The power laws are normalized at $\log M = -0.5$ ($0.31 M_\odot$).
The central parts of the cluster are clearly deficient in very low mass 
stars compared
to the local field population ($\alpha \approx -1.05$).  
\label{hyades_mf}}
\end{figure}

We interpret the absence of 
candidate Hyads of spectral types M9 through L4 as a lack
of free-floating objects in the
mass range $0.080 M_\odot > M > 0.060 M_\odot$, in
the central region of the cluster.   In Figure~\ref{hyades_mf},
we plot the RH99 mass function.  
This mass function is derived from the 112 sq. degree region
searched by Reid (1992, 1993); a few of the lowest mass objects
have been given a weight of 4 since they were found by other
searches covering only $\frac{1}{4}$ area.  Since our search covers only
28 sq. degrees, it is possible that we have simply been unlucky,
and that the brown dwarfs happen to lie in the other 84 sq. degrees.
If there were N very-low-mass objects in this part of the cluster,
the chance that we would find none is $(\frac{84}{112})^N$.
Thus we can rule out eight cluster objects in our mass range 
($0.08 - 0.06 M_\odot$, or 6.4 objects per $0.1\log M$) at the $90\%$
confidence limit.  We show this limit in Figure~\ref{hyades_mf}
as an arrow extending from the $90\%$ confidence limit down to
the observed value of 0.  This assumes that any differential mass segegration
effects are negligible over the two areas (the Reid M dwarfs are uniformly 
distributed over the 112 sq. degree area).  Even the larger Reid
(1992, 1993) area dioes not cover the entire cluster -- the tidal
radius is likely to be $\sim 10$ parsecs (\cite{hipphyades}).  

We define the mass function as:
$$\Psi(M) = \frac{dN}{dM} {\rm ~~~stars~ per ~unit ~mass}$$
$$\xi(M) = \frac{dN}{d \log (M)} {\rm ~~~stars ~per ~unit ~log(mass)}$$
The mass function is often fit to a power-law,
in which case $\Psi(M) \propto M^{-\alpha}$ or
$\xi(M) \propto M^{-\alpha+1}$.  

For nearby field stars, Reid \& Gizis (1997a) derived 
$\alpha = 1.05$, while Reid et al. (1999) have used
the 2MASS and Denis L dwarfs to estimate $\alpha \approx 1.3$ 
from  $0.1 M_\odot$ to $0.01 M_\odot$, although this estimate
has substantial uncertainties.  
Our result supports the conclusions of RH99 that the
central region of the Hyades is relatively deficient in very low mass dwarfs. 
A flat mass function in $\log M$ ($\alpha = 1$), for example, would 
require $\sim 20 $ objects in our mass range, or 5 over our limited 
search area.  In Figure~\ref{hyades_mf}, we plot three power-law mass
functions:  $\alpha = -1.05$ (nearby field objects), 
$\alpha = -0.7$ (\cite{rh99}), and $\alpha = 0$.
The data are suggestive that below $0.3 M_\odot$,
the mass function is even more depleted with $\alpha =0$.
Additional undiscovered binaries will tend to increase the mass function
at the lowest ends, but the failure of deep Keck imaging
(\cite{mac}) and HST snapshots (\cite{rg97b}) to find any 
brown dwarf secondaries in the Hyades, as well as the
general lack of brown dwarf companions (e.g., the Mayor et al. 1998 estimate
$\alpha \approx 0.4$ for brown dwarf secondaries), suggest that this 
is unlikely to be a large effect.  

The suggestion that the central region of the Hyades cluster is deficient in
very low mass dwarfs is supported by a comparison to the Pleiades.
B98 have compiled an estimate of
the Pleiades mass function from $7 M_\odot$ down to 
$0.045 M_\odot$.   The B98 mass function (their Figure 9) 
shows that the number of objects in the $\log M = -1.15$ 
bin (with width = 0.1 $\log M$) is about $80\%$ that of
the $\log M = -0.75$ bin.  The former bin is essentially
that of our survey (where we find 0 objects), while in
the latter bin Reid \& Hawley (1999) find 12 objects.
Alternatively, scaling from B98 and RH99 $\log M=-0.3$ region, 
we expect about 16 Hyades dwarfs in the $\log M = -1.15$ 
bin.  Thus 10-16 brown dwarfs in our sensitivity range are predicted 
in the Hyades using the Pleiades mass function (or 2.5-4 dwarfs in our
limited survey area -- to be compared with observed number of zero).  
Describing the Pleiades as a power law below $0.25 M_\odot$,
B98 estimate $\alpha =0.6 \pm 0.15$, 
while Martin et al. (1998) estimate a steeper $\alpha = 1.0 \pm 0.15$.
B98 note that this difference may not
be significant, since the two groups used different
corrections for completeness and cluster structure.  
Either is clearly not consistent with the RH99 mass function
extended with our data.  Since low luminosity secondaries
will be unresolved in both clusters, the issue of 
their contribution is not relevant to this comparison.

The deficiency in very low mass dwarfs may reflect either
a primordial difference in the Initial Mass Function (IMF) 
or the effect of dynamical evolution.  Perryman et al. (1998)
and Reid (1992) discuss evidence of mass segregation in the
cluster, with higher mass stars more concentrated towards the
cluster core.  Presumably the less concentrated 
very low mass dwarfs have been preferentially lost, or at
least moved to the outer parts of the cluster.  
On the other hand, Scalo (1998) has argued that there is
a strong indications of IMF variations in different clusters for  
the mass range $1-10 M_\odot$ with a spread of $\pm 0.5$ in $\alpha$.
Thus, the possibility that part of the difference is due to
the initial star formation cannot be ruled out.  

It should also be noted that this study only applies to brown dwarfs
near the center of the cluster, and therefore cannot distinguish
between the case where dynamical evolution has caused the 
very low mass dwarfs to be lost, or simply moved them to the outermost
parts of the cluster.  This survey could be extended to cover
more of the cluster, but considerable spectroscopic or astrometric 
follow-up would be necessary.  

An alternative interpretation is that brown dwarfs in the Hyades
are considerably fainter or cooler than models predict, and that therefore
they have eluded our survey.  While our knowledge of the exact mass range 
covered by our survey is dependent upon uncertain theoretical model
calculations, we would have detected the M9-L4 counterparts of 
the K2M field 2MASS discoveries.  Since the deficiency of M dwarfs members
according to RH99 is consistent with our lack of brown dwarfs,
we believe it is not necessary to invoke differences between 
Hyades and field L dwarfs to explain the paucity of L dwarf detections
relative to the Pleiades.  The question can only be resolved by
the discovery of Hyades L dwarfs, and 
the difference in age and composition of the
Hyades means that a detection of a Hyades L dwarf would be of considerable
interest, even if it tells us little about the IMF.

\section{Background Young Brown Dwarfs\label{background_bd}}

\begin{figure}
\plotone{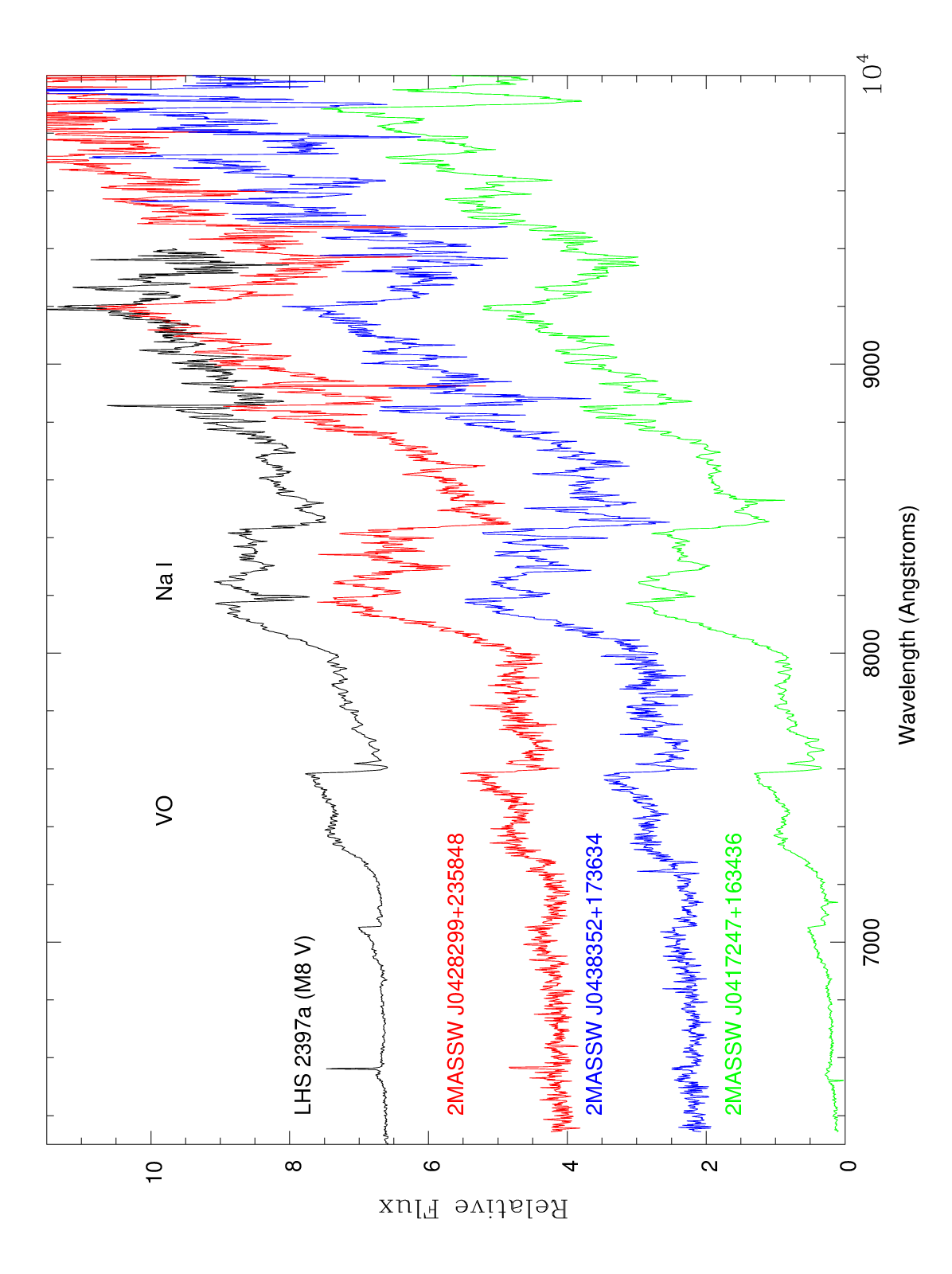}
%\plotone{fig_spectra.ps}
\caption{The spectra of three brown dwarf candidates compared to the field
M8 dwarf LHS 2397a.  These objects
have surface gravity features intermediate between field dwarfs and
giants.     
\label{figure-lowsg}}
\end{figure}

We plot the spectra of the three new 2MASS objects with unusual spectrscopic
properties in Figure~\ref{figure-lowsg}.  Compared to 
ordinary very late M dwarfs, these objects have weak Na at
8200 \AA, weak CaH, and strong VO absorption.  We measure the VO index
defined by Kirkpatrick et al. (1995) to have values of
$1.24 -1.3$ for the three Na-weak objects, while normal field
M8-M9 dwarfs observed with the same setup have VO values near 1.10 to 
1.18.  Spectral types for our objects are not well defined, since
there are no standards with similar weak features.  
We measure equivalent widths of $\sim 5$ \AA for the Na doublet
while normal field M dwarfs have values of $\sim 10$ \AA.  
(The definition of the pseudo-continuum and line limits is arbitrary
for these stars, and it is clear that our values are not 
comparable to those published by other authors).  

This behavior has been seen in five LH objects observed
spectroscopically by RH99, who interpret the Na-weak objects
as young brown dwarfs, probably associated with the Taurus-Auriga
association.  This identification is supported by the
astrometry of Harris et al. (1999), who find that the three LH Na-weak
objects which they measured lie at distances well beyond
the Hyades cluster.   The combination of weak surface gravity,
very low temperature, and high luminosity is consistent with the
young brown dwarf hypothesis.  We discuss the evidence for this
hypothesis in the contect of the new 2MASS objects below.  
The similarity of the 2MASS objects to the LH dwarfs suggests
that the 2MASS objects also lie at distances of $\sim 200$ parsecs.

The 2MASS spectra may be compared to both observations of
other very low mass dwarfs and theoretical models.  
The strength of the features immediately implies a surface 
gravity intermediate between that of M dwarfs and M giants.  
An additional constraint can be made by comparing to the
observations of confirmed Pleiades brown dwarfs by Martin et al. (1996).
Teide 1 and Calar 3 both show lithium absorption and a spectral type
of M8, implying they are brown dwarf at the age of the Pleiades
($\sim 125$ Myr, \cite{stauffer98}).   They have weak Na and
strong VO.  Specifically, Martin et al. (1996) show that the Na
doublet has an equivalent with about $60\%$ of that similar
field dwarfs.  The new 2MASS objects have even weaker Na features
(about $50\%$ the strength of field objects), which suggests
even lower surface gravities, and hence lower mass and younger age.

In order to compare to theoretical calculations, it is
necessary to estimate effective temperatures, luminosities, 
and gravities.  We have no parallax estimate, so we cannot
estimate the luminosity (although the low surface gravity
requires that the objects be above the main sequence; and the
Harris et al. parallaxes show that similar LH dwarfs do lie
above the main sequence.).  
We follow Luhman et al. (1997),
who estimate that PMS M8.5 dwarfs have $T_{eff} \approx 2600 K$. 
As for the surface gravity, we use the results of 
Schweitzer et al.'s (1996) modelling of the NaI and KI lines
in the M8 dwarf vB 10 (although their grainless models 
overestimate the temperature and 
are now considered obsolete, the relative comparisons of different
parameters should be useful).  Their Figure 2
indicates that the NaI line will decrease in strength 
by a factor of $\sim 2-3$ for a surface gravity change from the main
sequence value of $\log g=5.5$ to $\log g=4.5$.  
In contrast, their Figure 3 indicates that 
even increasing the metallicity of M8 dwarfs from 
$[m/H]=0.0$ to $[m/H]=0.5$ can not account for the
NaI weakening that we observe.  Since the observed
NaI line is weaker by at least a factor of 2, we estimate that
$4.4 < \log g < 4.7$.

\begin{figure}
\plotone{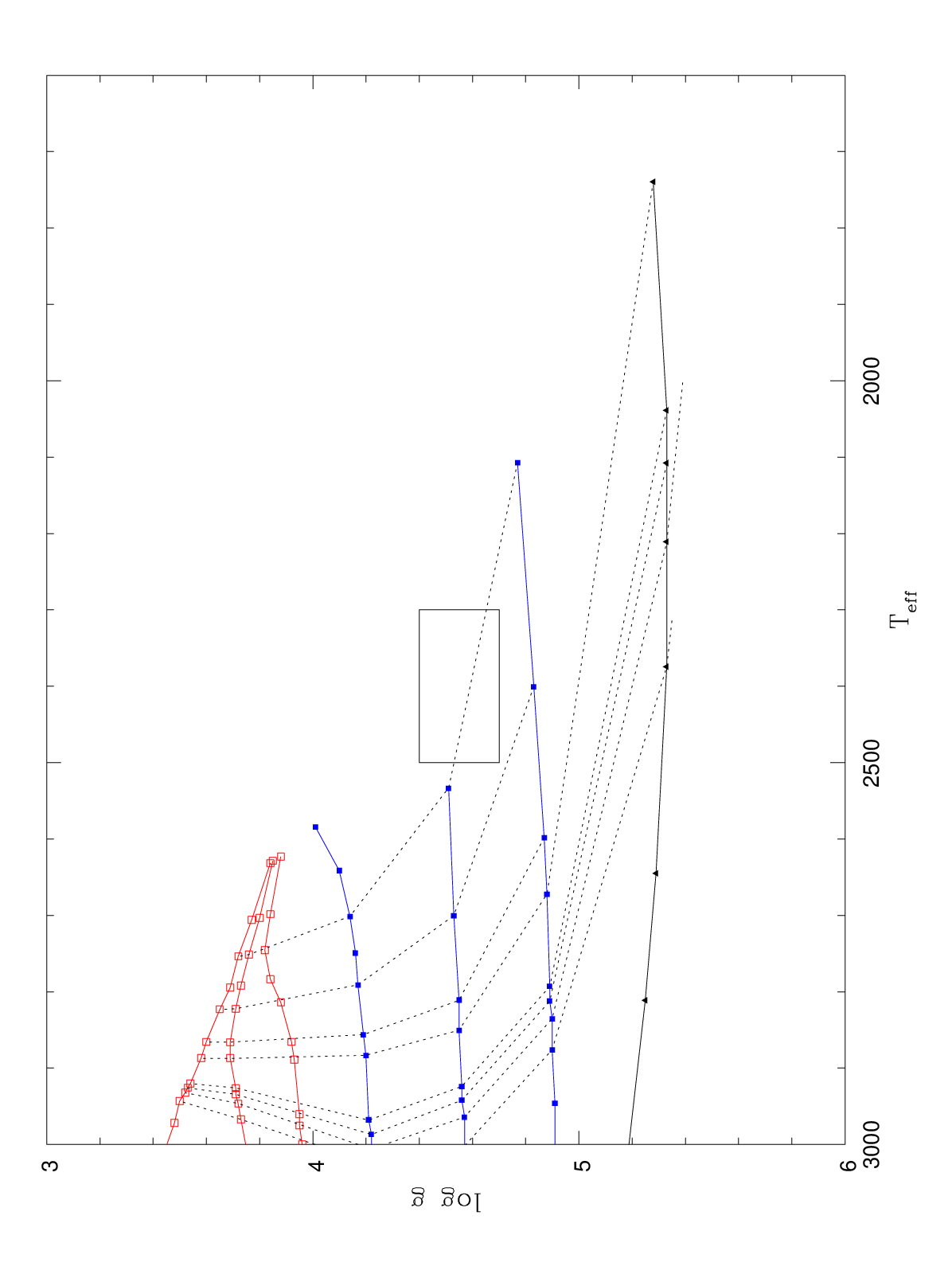}
%\plotone{fig_tg.ps}
\caption{\footnotesize Theoretical effective temperatures and gravities,
with the approximate region of the 2MASS low surface gravity objects
marked marked as the rectangle.  From top to bottom, the solid
lines indicate ages of $10^6$,$3 \times 10^6$,$5 \times 10^6$,
$10^7$,$3 \times 10^7$, $10^8$, and $10^9$ years.  From the left 
to the right,  the dashed lines represent the evolutionary tracks for 
$0.08$, $0.075$, $0.072$, $0.070$, $0.060$, $0.055$,
$0.045$, and $0.035 M_\odot$.  The models last three only extend to
$10^8$ years.  The 2MASS objects are consistent with ages of 
$\le 10^8$ years and substellar masses, a conclusion that is supported 
by the comparison of the spectra
to actual brown dwarfs in the Pleiades cluster.
\label{fig_tg}}
\end{figure}

Figure~\ref{fig_tg} shows model tracks from 
Baraffe \& Chabrier (priv. comm.).  The pre-main sequence 
models are preliminary,
and are an extension of the Baraffe et al. (1998) models.
The approximate position of the 2MASS
objects based on the above theoretical arguments is indicated.  
Masses of $lesssim 0.04 M_\odot$ and ages of
$\sim 30$ Myr are suggested by this comparison, if the models
may be extrapolated towards lower mass.  Since the
surface gravity estimate is at best a guess, and 
since the temperature scale of the models and/or
observations could be in error, these estimates must
be viewed with great caution.  Furthermore, these objects
are at the edge of the regime where dust formation, not yet
incorporated into the models, will become
important.  On the other hand, the theoretical estimates 
are consistent with the empirical comparison to the
observed Pleiades M8 brown dwarfs.  The luminosities predicted
by the models for masses $\lesssim 0.04 M_\odot$ and age
$\lesssim 30$ Myr are also consistent with the observed 
magnitudes of these objects at the distance  
the Taurus star formation region (150 pc. or more).

The chromospheric activity levels of the Na-weak objects show
a spread in activity.  2MASSW J0428299+235848, with an
equivalent width of $24$ \AA, is far more active than
the typical field M8-M9 dwarf.  2MASSW J0417247+163436,
in constrast, has an emission line of no more than $0.8$\AA.  
There are good reasons to expect that chromospheric
activity will {\it not} be a good age indicator for brown dwarfs:  
The M9.5 dwarf BRI 0021-0214 is a very rapid rotator 
but shows little activity (\cite{bm95}).  Tinney \& Reid (1998)
add the field brown dwarf LP 944-20 to this class of 
``inactive, rapid rotators'' and discuss the evidence that 
the activity-rotation has broken down at brown dwarf masses.  In this case, 
even very young brown dwarfs might not show strong chromospheric
activity.  Detailed studies of the differences between 
the active 2MASSW J0428299+235848 and the inactive 
2MASSW J0417247+163436 may shed light on this phenomenon.

The presence of young brown dwarfs at distances of $\approx
200$ parsecs, as noted by RH99, is consistent with the 
the presence of ROSAT-identified weak-lined T Tauri stars (WTTS).
Wichmann et al. (1996) find a surface density of 
about 1.6 WTTS per sq. degree in this region
and estimate ages of $\sim 10$ Myr for the WTTS.
Martin \& Magazzu (1999) classify many of these objects differently,
calling them post-T Tauri stars (PTTS) with ages up to $30 $ Myr.
Given the uncertainties, the 2MASS Na-Weak objects may well
be 10 Myr old, but older ages are also consistent with the
Martin \& Maguzzu analysis.   Combining the five RH99 young
brown dwarfs with our two ``central region'' objects, the surface 
density of  young brown dwarfs is $0.3$ per sq. degree.  
The order-of-magnitude equality of the two estimates is roughly
consistent with field stellar/substellar mass functions, but   
since the surveyed volumes, the target ages and masses, survey 
selection effects, and completeness are poorly understood for
both the WTTS/PTTS and 2MASS surveys more detailed comparison 
is not possible.  In particular, we may have excluded many similar
objects by cutting out the ``clump'' regions.  These regions of
high extinction are likely to  be associated with star formation,
and could be the source of the young brown dwarfs detected in
this paper.   It is also worth noting that both 2MASSW J0438352+173634
and 2MASSW J0428299+235848 lie near known T Tauri stars and 
CO emission regions (\cite{comap}; \cite{gomez}).  
 
It is interesting to note that we do not detect any L-type
low-surface gravity objects, which would correspond to 
even lower mass brown dwarfs.   Such objects may simply be too 
faint for 2MASS, or alternatively 
the Taurus association may not have formed brown dwarfs 
below $\sim 0.01 M_\odot$.

\section{Conclusions\label{conclusions}}

Our search for brown dwarfs has had mixed results -- no
brown dwarfs in our target cluster, yet younger brown dwarfs
in the background.  We show that in the central 28 sq. degrees
of the Hyades, there are no free-floating brown dwarfs down to 0.06 $M_\odot$.
This extends limits on the present-day mass function of
the Hyades, which is apparently declining for the lowest masses.

We identify a three of our late-M objects as young brown dwarfs
behind the Hyades cluster.  The weakening of the Na doublet
and the strengthening of the VO absorption is similar to,
but stronger than, the phenonemon observed in 
lithium brown dwarfs in the Plieades cluster.  
Model atmospheres also indicate that weak Na is due
to low surface gravity and not high metallicity.  
These objects therefore have low surface gravity
and therefore lie above the main sequence.  
Since models indicate that temperatures of very low mass objects 
decrease from 1 Myr to 1 Gyr, these three objects must be 
younger and lower mass than the field M8-M9 dwarfs (near $0.08 M_\odot$)
and Pleiades M8-M9 dwarfs (near $0.055 M_\odot$).  Similar
objects were reported by RH99.    
Many young stellar objects are known in this general direction,
so the presence of young brown dwarfs is not surprising.  

Use of additional 2MASS and POSSII data would allow the survey
to expanded to include the 112 degree area of the Hyades surveyed by Reid 
and the outermost parts of the cluster.  Even with a decreasing mass function
there is likely to be a Hyades L dwarf.  A side benefit will be uncovering 
additional examples of young, background brown dwarfs.

\acknowledgments

We thank the referee, John Stauffer, for his constructive comments.  
This work was funded in part by NASA grant AST-9317456
and JPL contract 960847.
This publication makes use of data products from the Two Micron All Sky 
Survey, which is a joint project of the
University of Massachusetts and the Infrared Processing and Analysis 
Center, funded by the National Aeronautics
and Space Administration and the National Science Foundation.
Some of the data presented herein were obtained at the W.M. Keck
Observatory, which is operated as a scientific partnership among 
the California Institute of Technology, the University of California 
and the National Aeronautics and Space
Administration.  The Observatory was made possible by the generous 
financial support of the W.M. Keck Foundation.
JEG accessed the Digitized Sky Survey images as a 
Guest User, Canadian Astronomy Data Centre, which is operated 
by the Herzberg Institute of Astrophysics, National Research Council 
of Canada. 
The Digitized Sky Surveys were produced at the Space Telescope Science 
Institute under U.S. Government grant NAG W-2166. The images of these 
surveys are based on photographic data obtained using the Oschin Schmidt 
Telescope on Palomar Mountain and the UK Schmidt Telescope. 
The plates were processed into
the present compressed digital form with the permission of these 
institutions.  The National Geographic Society - Palomar Observatory Sky 
Atlas (POSS-I) was made by the California Institute of Technology with 
grants from the National Geographic Society. 
The Second Palomar Observatory Sky Survey (POSS-II) was made by the 
California Institute of Technology with funds from the National Science 
Foundation, the National Geographic Society, the Sloan Foundation, 
the Samuel Oschin Foundation, and the Eastman Kodak Corporation. 
This research has made use of the Simbad database, operated at
CDS, Strasbourg, France.

%Tables look like this
%\begin{table}
%\dummytable\label{index}
%\end{table}

\begin{deluxetable}{ccccccccc}
\tablewidth{0pc}
\footnotesize
\tablenum{1}
\tablecaption{Known Objects}
\label{table-known}
\tablehead{
\colhead{Name} &
\colhead{RA (J2000)} &
\colhead{Dec}&
\colhead{J} &
\colhead{H} &
\colhead{K$_s$} & 
\colhead{B$_J$} &
\colhead{R$_F$} &
\colhead{I$_N$} 
}
\startdata
  Bry 262   & 04:21:49.57 & +19:29:08.9 & 12.66 &  12.02 & 11.66 & 21.5     & 17.1    & 14.9 \\ 
  Bry 804   & 04:38:27.71 & +16:00:11.0 & 13.33 &  12.70 & 12.38 & 22.3     & 16.7    & 15.4 \\   
  LH0416+14 & 04:19:36.98 & +14:33:32.9 & 14.36 &  13.65 & 13.22 &  \nodata & 20.7    & 16.0 \\
  LH0416+18 & 04:19:41.68 & +16:45:22.3 & 14.41 &  13.87 & 13.48 &  22.5    & 19.8    & 15.9 \\
  LH0418+13 & 04:20:50.16 & +13:45:53.2 & 14.22 &  13.55 & 13.01 &  \nodata & 20.3    & 16.7 \\
  LH0418+15 & 04:21:17.55 & +15:30:03.3 & 14.42 &  13.75 & 13.24 &  \nodata & 19.1    & 17.1 \\
  LH0419+15 & 04:22:30.80 & +15:26:31.4 & 14.36 &  13.53 & 13.04 &  \nodata & 19.9    & 17.1 \\
  LH0420+13 & 04:23:03.44 & +13:40:58.7 & 14.98 &  14.29 & 13.87 &  \nodata & 20.9    & 17.2 \\
  LH0420+15 & 04:23:24.74 & +15:41:45.7 & 14.59 &  14.08 & 13.71 &  \nodata & \nodata & 17.0 \\
  LH0424+15 & 04:27:39.32 & +15:07:34.5 & 13.35 &  12.79 & 12.49 &  22.0    & 18.0    & 15.5 \\
  LH0427+12 & 04:30:12.28 & +13:01:08.6 & 14.15 &  13.53 & 13.13 &  23.0    & 17.9    & 15.2 \\
\enddata
\end{deluxetable}

\begin{deluxetable}{lccccccccl}
\tablewidth{0pc}
\footnotesize
\tablenum{2}
\tablecaption{Candidates}
\label{table-targets}
\tablehead{
\colhead{Name} &
\colhead{RA (J2000)} &
\colhead{Dec}&
\colhead{J} &
\colhead{H} &
\colhead{K$_s$} & 
\colhead{R$_F$} &
\colhead{I$_N$} &
\colhead{H$\alpha$ EW} & 
\colhead{Sp. Type} 
}
\startdata
\cutinhead{Central Region}
2MASSWJ0422588+132937 & 04:22:58.84 & +13:29:37.3 & 14.41 & 13.65 & 13.35 & 19.5 & 16.2 & 8 &  M6.5  \\
2MASSWJ0419398+153659 &	04:19:39.84 & +15:36:59.5 & 14.22 & 13.41 & 13.15 & \nodata & \nodata & 0 & sdM?   \\
2MASSWJ0423242+155954 & 04:23:24.25 & +15:59:54.2 & 14.63 & 14.00 & 13.55 & 19.5 & 16.8 & 7\tablenotemark{a} & M5 \\
2MASSWJ0423242+155954\tablenotemark{b} & 04:23:24.21 & +15:59:53.8 & 14.71 & 14.02 & 13.53 \\
2MASSWJ0436223+164541 &	04:36:22.35 & +16:45:41.4 & 14.60 & 13.83 & 13.47 & 19.8 & 18.9 & 5 & M5   \\ 
LH0416+14 	      & 04:19:36.98 & +14:33:32.9 & 14.36 & 13.65 & 13.22 & 20.7 & 17.0 & 28.3 &  M7.5\tablenotemark{c,d} \\
2MASSWJ0432511+173009 & 04:32:51.19 & +17:30:09.2 & 14.70 & 14.01 & 13.55 & \nodata & 18.1 & \nodata & \nodata   \\
2MASSWJ0432511+173009\tablenotemark{b} & 04:32:51.20 & +17:30:09.5 & 14.71 & 14.05 & 13.54 \\
2MASSWJ0438443+155352 &	04:38:44.38 & +15:53:52.2 & 14.43 & 13.64 & 13.24 & 19.0 & 17.4 & 10 &  M6   \\
LH0418+13             & 04:20:50.16 & +13:45:53.2 & 14.22 & 13.55 & 13.01 & 20.3 & 16.7 & 12.6 & M8.5\tablenotemark{c}  \\
2MASSWJ0431420+162232 & 04:31:42.02 & +16:22:32.4 & 15.04 & 14.22 & 13.79 & 19.6 & 17.0 & 0 & reddened M \\
2MASSWJ0417247+163436 & 04:17:24.79 & +16:34:36.5 & 14.14 & 13.43 & 12.87 & 21.0 & 17.3 & 0 & M8\tablenotemark{d}  \\
2MASSWJ0431322+152620 & 04:31:32.22 & +15:26:20.5 & 15.43 & 14.56 & 14.11 & 20.6 & 17.2 & 0 & M4    \\
2MASSWJ0431500+152814 & 04:31:50.04 & +15:28:14.8 & 14.88 & 14.08 & 13.55 & 19.8 & 16.5  & 0 & M5   \\
2MASSWJ0431500+152814\tablenotemark{b} & 04:31:50.04 & +15:28:14.3 & 14.89 & 14.07 & 13.56 \\
2MASSWJ0430232+151436 & 04:30:23.26 & +15:14:36.8 & 15.72 & 14.60 & 14.25 & 20.4 & 16.8  & 0 & M5  \\
2MASSWJ0424045+143129 & 04:24:04.54 & +14:31:29.9 & 16.22 & 15.18 & 14.71 & 20.2 & 17.8  & 0 & non-M   \\
2MASSWJ0419434+143422 & 04:19:43.45 & +14:34:22.6 & 16.26 & 15.40 & 14.74 & 20.5 & 17.8  & \nodata & \nodata  \\
2MASSWJ0416582+140355 & 04:16:58.24 & +14:03:55.5 & 15.95 & 15.05 & 14.43 & 20.8 & 18.3  & 0 & M5  \\
2MASSWJ0434462+144802 & 04:34:46.21 & +14:48:02.7 & 16.14 & 15.29 & 14.61 & 20.3 & 19.0  & 0 & non-M   \\
2MASSWJ0435489+153719 & 04:35:48.99 & +15:37:19.5 & 16.37 & 15.55 & 14.80 & \nodata & 20.2  & 7 & M6   \\
2MASSWJ0438352+173634 & 04:38:35.20 & +17:36:34.5 & 16.53 & 15.45 & 14.79 & \nodata & \nodata  & 8 & M8\tablenotemark{d}  \\
\cutinhead{Other Sources}
2MASSWJ0421175+153003 &	04:21:17.55 & +15:30:03.3 & 14.42 & 13.75 & 13.24 & 19.1 & 17.1 &  6 & M6.5 \\ 
2MASSWJ0421521+151941 & 04:21:52.19 & +15:19:41.1 & 14.20 & 13.48 & 13.01  &19.1 &15.6 &  4 &  M6  \\
2MASSWJ0444539+181822 & 04:44:53.98 & +18:18:22.6 & 15.03 & 14.32 & 13.85  &19.2 &17.8 & 10 & M7  \\
2MASSWJ0443569+182824 &	04:43:56.90 & +18:28:24.8 & 15.43 & 14.65 & 14.08  & 20.5 &17.5& 5 & M6.5\\
2MASSWJ0428299+235848 &	04:28:29.99 & +23:58:48.4 & 16.16 & 15.31 & 14.64  & \nodata &\nodata& 24 & M8\tablenotemark{d}  \\
2MASSWJ0417456+212054 &	04:17:45.68 & +21:20:54.6 & 16.54 & 15.67 & 14.89  & 20.5 &18.3&0 &  M5 \\
2MASSWJ0426598+191233 &	04:26:59.88 & +19:12:33.1 & 16.96 & 15.87 & 14.89  & \nodata &19.3&0 & non-M \\
\enddata
\tablenotetext{a}{Value given for two averaged spectra.  First exposure had emission line with
$EW=14$ Angstroms; the second did not show emission.  Possibly a flare
occurred during the first observation.}
\tablenotetext{b}{Source lies in overlap region, both observations listed.}
\tablenotetext{c}{Spectral Information from Reid \& Hawley (1999).}
\tablenotetext{d}{Weak surface gravity features.}
\end{deluxetable}

\clearpage

%--------------------------BIBLIOGRAPHY---------------------------

\clearpage

\end{document}